\documentclass[twocolumn,preprintnumbers,amsmath,aps] {revtex4}
\usepackage{graphicx}
\usepackage{dcolumn}
\usepackage{bm}
\usepackage{color}
\usepackage{microtype}
\usepackage{multirow}
\usepackage{booktabs}
\begin{document}
\title{The superconducting energy gap in the hole-doped graphene beyond the Migdal's theory}


\author{Adam Z. Kaczmarek$^{1}$}\email{adam.kaczmarek@doktorant.ujd.edu.pl}
\author{Ewa A. Drzazga-Szcz{\c{e}}{\'s}niak$^{2}$}


\affiliation{$^1$Department of Theoretical Physics, Faculty of Science and Technology, Jan D{\l}ugosz University in Cz{\c{e}}stochowa, 13/15 Armii Krajowej Ave., 42200 Cz{\c{e}}stochowa, Poland}
\affiliation{$^2$Department of Physics, Faculty of Production Engineering and Materials Technology, Cz{\c{e}}stochowa University of Technology, 19 Armii Krajowej Ave., 42200 Cz{\c{e}}stochowa, Poland}

\date{\today}
\begin{abstract}
In this work we analyze impact of non-adiabatic effects on the superconducting energy gap in the hole-doped graphene. By using the Eliashberg formalism beyond the Migdal's theorem, we present that the non-adiabatic effects strongly influence the superconducting energy gap in the exemplary boron-doped graphene. In particular, the non-adiabatic effects, as represented by the first order vertex corrections to the electron-phonon interaction, supplement Coulomb depairing correlations and suppress the superconducting state. In summary, the obtained results confirm previous studies on superconductivity in two-dimensional materials and show that the corresponding superconducting phase may be notably affected by the non-adiabatic effects. 
\end{abstract}
\maketitle
\vspace{0.5cm}

\section{Introduction}

The discovery of graphene has led to the ever-growing interested in its electronic properties \cite{castro2009}. Among various electronic aspects, a notable attention was given to the induction of the conventional superconducting state in this material. In this respect, one the most promising scenarios were realized via doping graphene with a foreign atoms \cite{profeta2012,salvini2010,zhou2015,gholami2018}. In general, there are two main routes to enhance graphene's electron-phonon coupling ($\lambda$). First is the so-called surface functionalization, where metal atoms are deposited at the surface of the monolayer \cite{mcchesney2010,profeta2012,zhou2014}. Unfortunately, in this approach, the resulting critical temperature of the superconductive state ($T_C$) is rather low. However, the second strategy aims at the introduction of impurities which act like a electron or hole dopants and lead to the much higher $T_C$ values \cite{salvini2010,zhou2015,szczesniak2021}. 

Along with a relatively high $T_C$ values of the substitutionally doped graphene, such material exhibits shallow conduction band \cite{salvini2010,zhou2015}, similarly to fullerenes and fullerides \cite{dean2010,cao2018,pietronero1992,gunnarsson1997}. This leads to the significant value of the phonon and electron energy scales ratio ($\omega_D/E_F$, where $\omega_D$ is the Debye's frequency and $E_F$ denotes Fermi energy), which cannot be neglected in the framework of the Migdal's theorem \cite{migdal1958}. Such behavior results in the non-adiabatic effects strongly influencing the superconducting phase \cite{pietronero1995,grimaldi1995}. As suggested in \cite{pietronero1995, pietronero2006}, the non-adiabatic effects may have nontrivial impact on the electron-phonon interaction, that can be observed based on the order parameter behavior. For example, proper characterization of these effects in graphene was recently described for the electron-doped graphene structures in \cite{szczesniak2021}. To be specific, the Authors of \cite{szczesniak2021} have shown that the contribution of the non-adiabatic effects is rising upon the increase of the Coulomb interaction.

With respect to the above, we investigate the non-adiabatic effects in case of the hole-doped graphene, to determine their impact on the order parameter and the $T_C$ value. To do so, we employ the Eliashberg equations \cite{eliashberg1960} with the first order vertex-corrections  \cite{pietronero1992,pietronero1992b,grimaldi1995}. The calculations are done for the $50 \%$ boron-doped graphene structure ($h$-CB) under biaxial tensile strain $\epsilon =5\%$ and at the moderate level of the dopant electrons ($n=-0.2 |e|/\text{unit cell}$) \cite{zhou2015}.

\section{Theoretical model}

As already mentioned, the present analysis is based on the Eliashberg formalism \cite{eliashberg1960, carbotte1990}. Conventionally this formalism is employed within the adiabatic regime {\it i.e.} by assuming the Migdal's theorem \cite{migdal1958}. However, to analyze the non-adiabatic effect the Eliashberg equations are additionally generalized here by considering the first order vertex corrections to the electron-phonon interaction \cite{pietronero1995, pietronero1992b, freericks1997}.

Specifically, we assume that the adiabatic Eliashberg equations on the imaginary axis have the form:
\begin{align}
\label{eq1}
    \phi_{n}=\pi k_{B}T\sum^{M}_{m=-M}
\frac{[K_{n,m}-\mu^{\star}\theta(\omega_{c}-|\omega_{m}|)]}{\sqrt{\omega^{2}_{m}Z_{m}^{2}+\phi^{2}_{m}}}\phi_{m},
\end{align}
\begin{align}
\label{eq2}
Z_{n}=1+\pi k_{B}T\sum^{M}_{m=-M}
\frac{K_{n,m}}{{\sqrt{\Delta_{m}^2+\omega^{2}_{m}}}}\frac{\omega_{m}}{\omega_{n}}Z_{m}.
\end{align}
where $\phi_n=\phi(i\omega_n)$ denotes the order parameter function and $Z_n=Z(i\omega_n)$ is the renormalization factor of the wave function. In what follows, $k_{B}$ is the Boltzmann constant, $T$ denotes the temperature, and $\omega_{n}$ represents the $n$-th Matsubara frequency ($\omega_{n}=\pi k_{B}T\left(2n+1\right)$). In this framework, $M$ denotes the cut-off value for the calculations and is equal to $1100$, so the numerical calculations are stable for $T>5$ K. Moreover, $\mu_{n}^{\star}=\mu^{\star}\theta \left(\omega_{c}-|\omega_{n}|\right)$ is the Coulomb pseudopotential which models the depairing correlations; where $\theta$ is the Heaviside function and $\omega_{c}$ represents the cut-off frequency.

In the above equations, the electron-phonon pairing kernel is expressed as:
\begin{align}
\label{eqS03}
  K_{n,m} \equiv 2 \int_0^{\omega_D} \text{d}\omega \frac{\omega}{4\pi^2k_B^2 T^2(n-m)^2+\omega^2}\alpha^2 F(\omega), 
\end{align}
where $\alpha^2 F(\omega)$ denotes the Eliashberg function for a given $\omega$ phonon energy:
\begin{equation}
\label{eqS04}
\alpha^2F(\omega) = {1\over 2\pi \rho\left(E_{F}\right)}\sum_{{\bf q}\nu} 
                    \delta(\omega-\omega_{{\bf q}\nu})
                    {\gamma_{{\bf q}\nu}\over\omega_{{\bf q}\nu}},
\end{equation}
whereas:
\begin{equation}
\begin{aligned}
\label{eqS05}
\gamma_{{\bf q}\nu}=2\pi\omega_{{\bf q}\nu} \sum_{ij}
                \int {d^3k\over \Omega_{BZ}}  |g_{{\bf q}\nu}({\bf k},i,j)|^2  \\
 			\times\delta(E_{{\bf q},i}-E_{F})\delta(E_{{\bf k}+{\bf q},j}-E_{F}).
\end{aligned}
\end{equation}
In Eq. (\ref{eqS05}), the $\omega_{{\bf q}\nu}$ gives values of the phonon energies and $\gamma_{{\bf q}\nu}$ denotes the phonon linewidth. In this context, the electron-phonon coefficients are represented by $g_{{\bf q}\nu}({\bf k},i,j)$ and $E_{{\bf k},i}$ stands for the electron band energy. Note that the higher order corrections are not included in Eq. (\ref{eqS03}), and that the momentum dependence of electron-phonon matrix elements has been neglected in Eq. (\ref{eqS01}) and Eq. (\ref{eqS02}) (in accordance to the local approximation). Therefore, the order parameter can be written as: $\Delta_{n}(T, \mu^*)=\phi_{n}/Z_{n}$. Finally, we note that for the purpose of our research, we use Eliashberg function given in \cite{zhou2015}. It is important to remark, that the resulting cutoff frequency in Eq. (\ref{eq1}) is $\omega_C=10 \omega_{\text{max}}$ with the maximum phonon frequency equal to $\omega_{\text{max}}=124.47$ meV.

With respect to the presented adiabatic equations, the introduction of the first-order vertex correction terms leads to the non-adiabatic Eliashberg equations (N-E) of the following form \cite{freericks1997,szczesniak2021}:
\begin{widetext}
\begin{eqnarray}
\label{eqS01}
\nonumber
\phi_{n}&=&\pi k_{B}T\sum_{m=-M}^{M}
\frac{K_{n,m}-\mu_{m}^{\star}}
{\sqrt{\omega_m^2Z^{2}_{m}+\phi^{2}_{m}}}\phi_{m} - \beta\frac{\pi^{3}\left(k_{B}T\right)^{2}}{4E_{F}}\\ \nonumber
&\times&\sum_{m=-M}^{M}\sum_{m'=-M}^{M}
\frac{K_{n,m}K_{n,m'}}
{\sqrt{\left(\omega_m^2Z^{2}_{m}+\phi^{2}_{m}\right)
       \left(\omega_{m'}^2Z^{2}_{m'}+\phi^{2}_{m'}\right)
       \left(\omega_{-n+m+m'}^2Z^{2}_{-n+m+m'}+\phi^{2}_{-n+m+m'}\right)}}\\
&\times&
\left(
\phi_{m}\phi_{m'}\phi_{-n+m+m'}+2\phi_{m}\omega_{m'}Z_{m'}\omega_{-n+m+m'}Z_{-n+m+m'}-\omega_{m}Z_{m}\omega_{m'}Z_{m'}
\phi_{-n+m+m'}
\right),
\end{eqnarray}
and
\begin{eqnarray}
\label{eqS02}
\nonumber
Z_{n}&=&1+\frac{\pi k_{B}T}{\omega_{n}}\sum_{m=-M}^{M}
\frac{K_{n,m}}{\sqrt{\omega_m^2Z^{2}_{m}+\phi^{2}_{m}}}\omega_{m}Z_{m} - \beta\frac{\pi^{3}\left(k_{B}T\right)^{2}}{4E_{F}\omega_{n}}\\ \nonumber
&\times&\sum_{m=-M}^{M}\sum_{m'=-M}^{M}
\frac{K_{n,m}K_{n,m'}}
{\sqrt{\left(\omega_m^2Z^{2}_{m}+\phi^{2}_{m}\right)
       \left(\omega_{m'}^2Z^{2}_{m'}+\phi^{2}_{m'}\right)
       \left(\omega_{-n+m+m'}^2Z^{2}_{-n+m+m'}+\phi^{2}_{-n+m+m'}\right)}}\\
&\times&
\left(
\omega_{m}Z_{m}\omega_{m'}Z_{m'}\omega_{-n+m+m'}Z_{-n+m+m'}+2\omega_{m}Z_{m}\phi_{m'}\phi_{-n+m+m'}-\phi_{m}\phi_{m'}\omega_{-n+m+m'}Z_{-n+m+m'}
\right).
\end{eqnarray}
\end{widetext}
Note that when vertex-corrections contribution terms are neglected, the above Eliashberg equations take the adiabatic form of Eqs. (\ref{eq1}) and (\ref{eq2}).

\section{Results and discussion}

\begin{center}
\begin{figure*}[!t]
\includegraphics[width=170 mm]{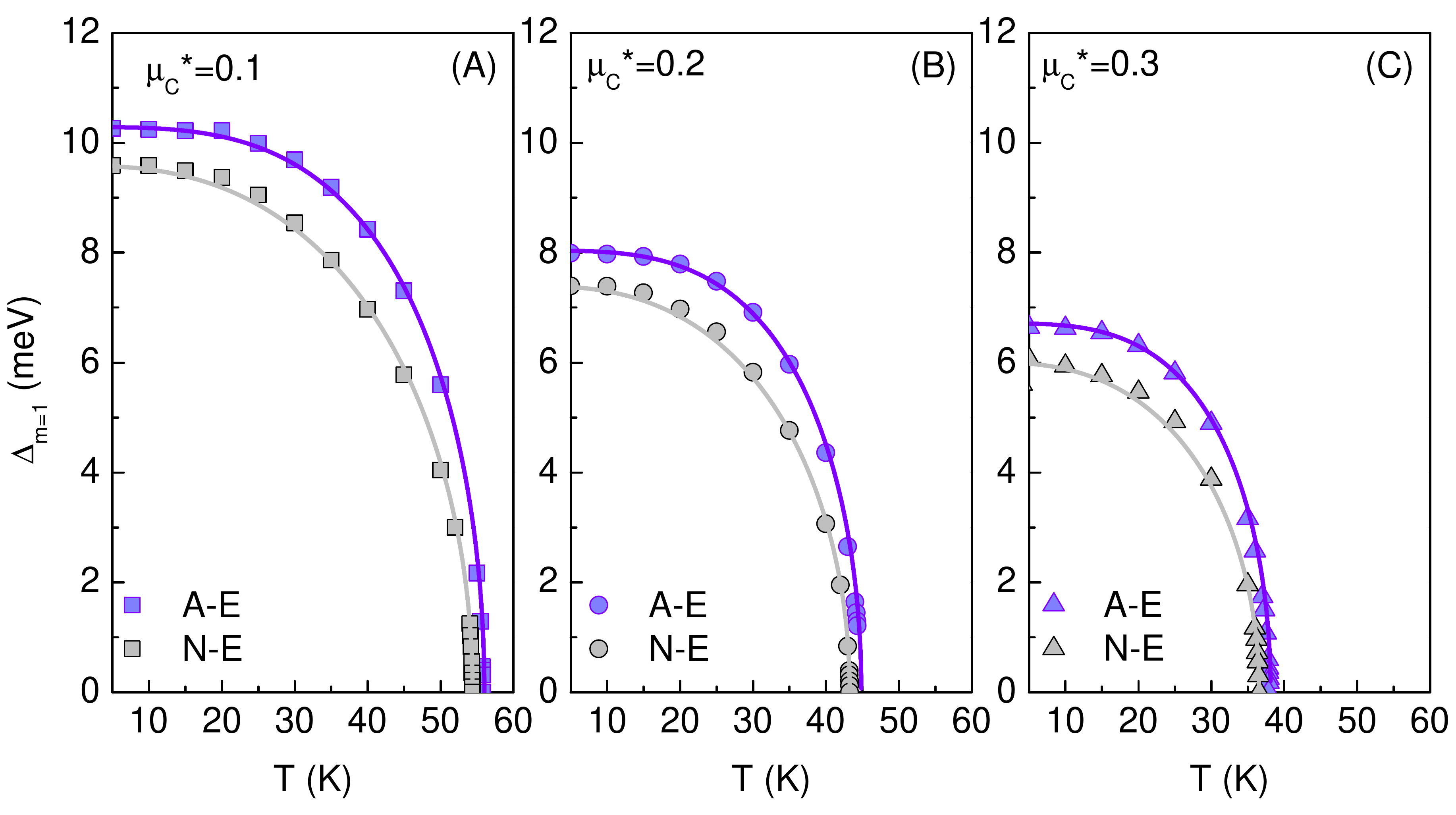}
\caption{The temperature dependence of the order parameter for selected $\mu$ values. The adiabatic Eliashberg solutions are marked by the purple symbols whereas the non-adiabatic results by the gray ones. The solid lines are guide to an eye.}
\label{f1}
\end{figure*}
\end{center}

We begin our discussion by noting that the adiabatic (A-E) and the non-adiabatic (N-E) equations presented in the previous section allows us to obtain the order parameter dependence on the temperature in the form: $\Delta_n(T,\mu^*)=\frac{\phi_n}{Z_n}$. This is done by using the numerical techniques presented originally in \cite{freericks1997,Szczesniak2006,szczesniak2019}. In such analysis, the special attention is paid to the maximum value of the order parameter $\Delta_{m=1}(T,\mu^*)$, which equals to zero when $T=T_C$ and $\mu^*=\mu^*_C$.  In other words, it allows us to determine the critical value of the temperature for a given critical value of the Coulomb pseudopotential ($\mu^*_C$), which is considered here as a free parameter. The latter one is assumed due to the fact that there are no experimental predictions of the $T_C$ for the hole-doped graphene in the literature.

Specifically, our analysis is constrained to the three different values of $\mu^*_C$. In this way we can span relatively wide range of the $\mu^*$ values, allowing for future comparisons with existing literature on the graphene-based superconductors \cite{szczesniak2009, szczesniak2019, szczesniak2015} or with the experimental estimates. Fig. (\ref{f1}) depicts results of the numerical analysis for three different values of $\mu^*$. The adiabatic solutions are represented by purple symbols, while the gray ones corresponds to the non-adiabatic Eliashberg equations associated with the vertex corrections. The presented results exhibits conventional behavior for the superconductors with the electron-phonon pairing mechanism, where $\Delta_{m=1}$ has plateau at lower temperatures and decreases quickly for the higher ones.

However, the main observation can be made when comparing the adiabatic and non-adiabatic results. Namely, the inclusion of the vertex corrections in the theoretical framework leads to the decrease of the $\Delta_{m=1}$ parameter for the entire range of $T$ and $\mu^*$. The thermodynamic properties originated from this fact may be observed in the experiment. To be specific, lower $\Delta_{m=1}$ for the non-adiabatic equations leads to $T_C \in \left< 54.4,36.6 \right>$ K in comparison with the values obtained for the adiabatic scenario: $T_C \in \left< 55.8,37.9 \right>$ K. Therefore, the non-adiabatic effects slightly lower critical temperature by $\sim 1 \%$. We note that upon comparison with the electron-doped graphene \cite{szczesniak2021}, the decrease of the $T_C$ is smaller. In fact, for the $h-$CN structure, changes are noticeable. To be specific, in the non-adiabatic framework the critical temperature is decreasing by $\sim 30\%$. Henceforth, the non-adiabatic effects in the hole-doped graphene are more favorable from the standpoint of keeping nominal $T_C$ as high as possible. Influence of the non-adiabatic effects are another suppressor of the high $T_C$ values besides the Coulomb pseudopotential. Implications of the non-adiabatic effects can also be seen at the origin of the temperature axis. In particular, the $\Delta_{m=1}$ for $T_0$, which corresponds to the half-width of the superconducting gap ($\Delta$), is also lower for the non-adiabatic solutions than in the adiabatic case. Upon increase of the $\mu^*$, $\Delta \in \left< 10.3,6.7 \right>$ meV for the adiabatic case and $\Delta \in \left< 9.6,6.0 \right>$ meV for the non-adiabatic case. Hence, the inclusion of the vertex corrections decrease the value of $\Delta$ by $\sim 5\%$ and $\sim 10\%$, respectively. By comparison with the twin $h-$CN material, the result of these corrections are significantly smaller \cite{szczesniak2021}. In fact, the non-adiabatic effects in the electron-doped graphene are decreasing value of the order parameter by $ \sim 40 \%$.

\begin{table*}
\centering
\caption{The thermodynamic quantities of the hole-doped graphene, as calculated in the present paper: the critical temperature $T_C$, the superconducting gap half-width $(\Delta)$ and the characteristic ratio $R_{\Delta}$. Results are obtained for the adiabatic (A-E) and non-adiabatic (N-E) Eliashberg approach. }
\begin{tabular*}{\textwidth}{@{\extracolsep{\stretch{1}}}*{8}{l}@{}}
\toprule
\hline
$\mu^{\star}$ & $T_{C}$ (A-E) K & $T_{C}$ (N-E) K & $\Delta$ (A-E) meV & $\Delta$ (N-E) meV & $R_{\Delta}$ (A-E) & $R_{\Delta}$ (N-E) \\
\hline
\midrule
0.1 & 55.8 & 54.4 & 10.3 & 9.6 & 4.27 & 4.08 \\
0.2 & 44.1 & 43.1 & 8.1  & 7.3 & 4.16 & 3.96 \\
0.3 & 37.9 & 36.6 & 6.7  & 6.0 & 4.09 & 3.79 \\
\hline
 \bottomrule                             
\end{tabular*}
\label{tab}
\end{table*}

From the perspective of the future experimental search, obtained values of the $\Delta$ and $T_C$ parameters may not be sufficient for the identification of the non-adiabatic effects in the hole-doped graphene. Thus, one should calculate characteristic ratio for the order parameter \cite{carbotte1990}:
\begin{align}
R_{\Delta}\equiv2\Delta(0)/k_{B}T_{C}.
\label{eq8}
\end{align}
The Eq. (\ref{eq8}) originates from the BCS theory \cite{bardeen1957,bardeen1957b} and as a dimensionless parameter it is important from the perspective of experiments conducted in the future. 
Here, by using Eq. (\ref{eq8}) we obtain $R_\Delta \in \left< 4.08,3.79 \right>$ and $R_\Delta \in \left< 4.28,4.09 \right>$. Again, the non-adiabatic effects lead to the reduction of the thermodynamic parameter value. It is important to note that for both types of the Eliashberg equations, values of parameter $R_\Delta$ are higher than the standard BCS value of $3.53$ \cite{carbotte1990,bardeen1957,bardeen1957b}. Moreover, the difference between the non-adiabatic and adiabatic values of the characteristic ratio $R_\Delta$ is much smaller than the ratio encountered in the case of its nitrogen-doped counterpart \cite{szczesniak2021}. From the analysis presented above, one can also conclude that the retardation effects and strong coupling have an impact on the superconducting state in the hole-doped graphene.

\section{Summary}

We have tackled theoretical and numerical analysis within the Eliashberg theory to discuss possible impact of the non-adiabatic effects on the thermodynamic properties of the superconducting state in the hole-doped graphene ($h$-CB). Our analysis has been performed to analyze behavior of the critical temperature ($T_C$), the superconducting gap half-width ($\Delta$) and the dimensionless BCS-ratio for the order parameter ($R_{\Delta}$). Values of these parameters for the adiabatic and non-adiabatic equations are summarized and presented in Table (\ref{tab}). Is it clear, that inclusion of the non-adiabatic effects, via vertex corrections to the electron-phonon interaction, reduces values of the thermodynamic parameters. It is also worth to notice that for the stronger electron-coupling displayed by the higher $\mu^*$ values, the non-adiabatic effects become slightly stronger. In other words, the Coulomb interaction is supplemented by the non-adiabatic effects. Moreover, these effects are significantly smaller than in the case of the electron-doped graphene analyzed in \cite{szczesniak2021}. It means that the hole-doped graphene is more robust against the non-adiabatic effects, since these effects are decreasing its critical temperature ($T_C$) minimally.

Finally, the results presented here supplement observations conducted for the electron-doped graphene structure \cite{szczesniak2021}. In the comparison with the electron-doping, the hole-doped structure is more robust against the non-adiabatic effects \cite{szczesniak2021}. However, the superconducting properties will still be decreased in the framework of the  vertex-corrected Eliashberg equations. In general, the hole-doped graphene may be a still interesting choice for the phonon-induced superconducting material.

\bibliographystyle{apsrev}
\bibliography{bibliography}

\end{document}